\documentclass[fleqn,11pt]{wlscirep}

\usepackage{times}
\usepackage{authblk}
\usepackage{xr}
\usepackage[utf8]{inputenc}
\usepackage[T1]{fontenc}
\usepackage{mathtools}
\usepackage{enumitem}
\usepackage[scr=rsfs]{mathalpha}
\usepackage{mathtools}
\usepackage{amsmath,amsthm,amssymb,amsfonts}
\usepackage{color}
\usepackage{color,soul}
\usepackage{graphicx}
\usepackage{lipsum}
\usepackage{float}
\usepackage{verbatim}
\usepackage[mathlines]{lineno}
\usepackage{physics}
\usepackage{algorithm,algpseudocode}
\usepackage{siunitx}
\usepackage[normalem]{ulem}
\usepackage{lineno}

\DeclareSIUnit\Molar{\textsc{M}}

\algnewcommand{\Inputs}[1]{%
  \State \textbf{Inputs:}
  \Statex \hspace*{\algorithmicindent}\parbox[t]{.8\linewidth}{\raggedright #1}
}
\algnewcommand{\Initialize}[1]{%
  \State \textbf{Initialize:}
  \Statex \hspace*{\algorithmicindent}\parbox[t]{.8\linewidth}{\raggedright #1}
}

\title{Dynamic cluster field modeling of collective chemotaxis}

\author[1]{Aditya Paspunurwar}
\author[2]{Adrian Moure}
\author[1,3,4,*]{Hector Gomez}

\affil[1]{School of Mechanical Engineering, Purdue University, 585 Purdue Mall, West Lafayette, 47907, IN, USA}
\affil[2]{Department of Mechanical and Civil Engineering, California Institute of Technology, 1200 East California Boulevard, Pasadena, 91125, CA, USA}
\affil[3]{Weldon School of Biomedical Engineering, Purdue University, 206 S. Martin Jischke Drive, West Lafayette, 47907, IN, USA}
\affil[4]{Purdue Institute for Cancer Research, Purdue University, 201 S. University Street, West Lafayette, 47907, IN, USA}
\affil[*]{Corresponding author: hectorgomez@purdue.edu}


\vspace{-1em}

\begin{abstract}
Collective migration of eukaryotic cells is often guided by chemotaxis, and is critical in several biological processes, such as cancer metastasis, wound healing, and embryogenesis. Understanding collective chemotaxis has challenged experimental, theoretical and computational scientists because cells can sense very small chemoattractant gradients that are tightly controlled by cell-cell interactions and the regulation of the chemoattractant distribution by the cells. Computational models of collective cell migration that offer a high-fidelity resolution of the cell motion and chemoattractant dynamics in the extracellular space have been limited to a small number of cells. Here, we present Dynamic cluster field modeling (DCF), a novel computational method that enables simulations of collective chemotaxis of cellular systems with $\mathcal{O}(1000)$ cells and high-resolution transport dynamics of the chemoattractant in the time-evolving extracellular space. We illustrate the efficiency and predictive capabilities of our approach by comparing our numerical simulations with experiments in multiple scenarios that involve chemoattractant secretion and uptake by the migrating cells, regulation of the attractant distribution by cell motion, and interactions of the chemoattractant with an enzyme. The proposed algorithm opens new opportunities to address outstanding problems that involve collective cell migration in the central nervous system, immune response and cancer metastasis.
\end{abstract}

\begin{document} 
\flushbottom
\maketitle
\keywordname{ Collective cell migration, Chemotaxis, Phase field modeling, Clustering, Reallocation}

\section*{Introduction}


From the smallest microorganisms to the most intricate multicellular life forms, chemotaxis emerges as a universal orchestrator, guiding cellular migration by the gradient of a chemical compound.
In particular, eukaryotic cells possess an exceptional ability to interpret and respond to the most subtle chemical gradients \cite{zigmond1977ability,parent1999cell}.
Unveiling the mechanisms of chemotaxis requires an in-depth understanding of the cell's response to a chemical gradient as well as detailed knowledge of the chemical gradient that is presented to the cell under different circumstances.
However, the study of chemotaxis has been dominated
by the notion of cells following a chemotactic gradient in an environment where chemoattractants diffuse both in the intracellular and extracellular regions \cite{lushnikov2008macroscopic, zhao2017dynamic,palsson2001three}.
This is insufficient because as cells sense chemicals and move through them, they modify the spatial distribution of the chemoattractant in various ways.
For example, some cells secrete chemoattractant; some cells uptake chemoattractant through their membrane receptors, and the motion of any cell changes the extracellular space continuously where the chemoattractant diffuses and reacts with other chemicals.
The action of cells on the chemoattractant distribution is small in the migration of single cells, but it becomes critical in collective cell migration \cite{dona2013directional}.
An example of particular significance is the chemotactic migration of non-confluent cell groups which is critical in embryogenesis, immune response, and metastasis, but remains poorly understood.
Unveiling the mechanism of chemotaxis of non-confluent cell groups poses challenges to existing experimental and computational techniques.
Despite techniques to measure cAMP/Ca$^{2+}$ gradients exist\cite{howze2022improving,annamdevula2020measurement}, we lack tools to visualize the activity of a wide range of extracellular guiding signals from various chemoattractant \cite{dona2013directional}. 
Collective chemotaxis has been explored computationally in various scenarios including angiogenesis \cite{heck2015computational} and vasculogenesis \cite{boas2018cellular}. 
Computational efforts that use high-fidelity models of chemotaxis have emphasized the study of individual cells \cite{song2010three,moure2018three}.
In this simplified scenario, computer simulations achieved substantial success in predicting the two-way interaction between the cell and chemoattractant.
One noteworthy category of models is that of phase-field methods \cite{levine2006directional,gomez2017computational,chen2002phase,emmerich2003diffuse}, which have been used to simulate the uptake of chemoattractant by the cell membrane and the transport dynamics of the chemoattractant in the time-evolving extracellular space.
In the phase-field method, the location and shape of a cell moving on a planar substrate are implicitly defined by a scalar function $\phi$ (the phase-field) which takes the value $\phi \approx 1$ inside the cell and the value $\phi \approx 0$ outside the cell.
The phase field is the solution to a partial differential equation (PDE) on a fixed domain and transitions smoothly between $0$ and $1$, identifying not only the intracellular and extracellular spaces but also the cell's membrane.
Thus, the phase-field method allows to incorporate classical models of membrane mechanics, uptake or secretion of chemoattractant at the cell's membrane and offers a simple way to compute chemoattractant transport dynamics in the time-evolving extracellular space avoiding the complication of moving meshes or another type of moving boundary technique.
However, efforts to extend these high-fidelity models to collective cell migration have been tempered by computational costs because predicting the motion of $N$ cells requires solving at least $N$ coupled PDEs, which has only been done in the literature for $N\sim 15$; see \cite{moure2021phase}.
%
%
%
%
Researchers have often resorted to either homogenized phenomenological models which do not resolve the individual position of the cells but give an estimate of the average cell density \cite{dallon1997discrete},
or a discrete cell model that solves chemoattractant dynamics across the entire domain and then interpolates to the cell's membrane to update chemotactic motion \cite{palsson2001three, palsson2009camp}.
Although these models have been successful in predicting cell migration patterns of more than {10,000} cells, they do not restrict the reaction and diffusion of chemoattractant in the extracellular space which may hinder our ability to provide a mechanistic understanding of collective chemotaxis, such as displacement and accumulation of chemoattractant influenced by cell movement.
Here, we introduce dynamic cluster field (DCF) modeling, a new computational method capable of predicting the chemotactic migration of $N=\mathcal{O}(1000)$ cells accounting for cell-cell and cell-chemoattractant interactions while accurately solving the chemoattractant transport dynamics in the extracellular space whose geometry and topology change with time. \\


The DCF approach uses fundamental properties of the phase-field theory to solve the transport dynamics of the chemoattractant on the extracellular space and specifically designed algorithms to dynamically allocate and de-allocate multiple cells to a \emph{cluster field}, thus reducing the number of PDEs to be solved and enabling us to achieve a $60$-fold increase in the number of simulated cells compared to existing algorithms that use the same level of fidelity. \\


We showcase the DCF method by simulating a wide range of scenarios, including co-attraction of cells that secrete chemoattractant, migration of cells that generate attractant gradients by themselves due to uptake of a guiding chemical compound, and systems that involve the interaction of a chemoattractant with an enzyme on the extracellular space.
Our simulations are in good agreement with experiments and illustrate that an accurate solution of chemoattractant dynamics in the extracellular space is essential to capture the complexity of collective chemotaxis.



\section*{Results}

\subsection*{\textbf{Model description}}

\begin{figure}[h]
	\centering
	\includegraphics[width=\linewidth]{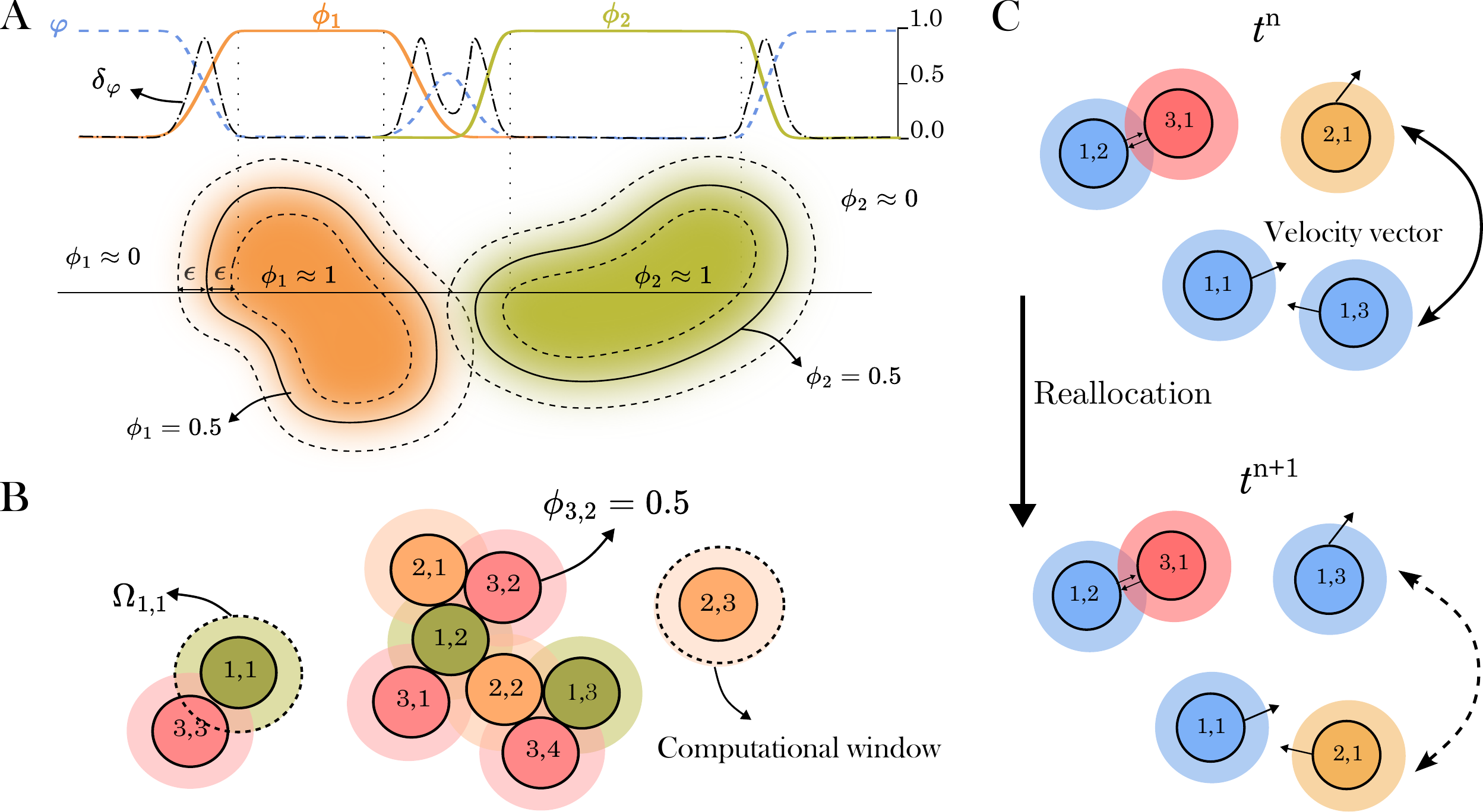}
	\caption{ {\small \textbf{(A) Diffuse-interface approach.} 
    We localize cell $\alpha$ and the extracellular environment with the phase field $\phi_\alpha$ and the marker $\varphi = 1 - \sum_{\alpha} \phi_\alpha$, respectively, which take the value $1$ inside the corresponding region and $0$ outside. 
    We localize the cells membrane with the marker $\delta_\varphi$ (see the main text), which is maximum at $\phi_\alpha = 0.5$ and decreases to $0$ as $\phi_\alpha$ approaches $0$ and $1$. 
    \textbf{(B) Clustering strategy.} We group multiple distant cells by assigning them to a single cluster field $\mathbb{C}_i=\sum_j\phi_{i,j}$, where index $j$ identifies cells within cluster $i$.
    Cells with the same color in the figure belong to the same cluster.
    The shaded region surrounding each cell $(i,j)$ represents the computational window $\Omega_{i,j}$, where Eq.~\eqref{Eqn:phase-field} is solved.
    \textbf{(C) Reallocation strategy.} When two cells in the same cluster approach (cells $(1,1)$ and $(1,3)$ in the figure), we switch the allocation of one of them (cell $(1,3)$) with a cell from a different cluster (cell $(2,1)$) such that the computational windows of cells in the same cluster do not overlap after reallocation (note that cell $(3,1)$ cannot be transferred to cluster 1).
    }    }
	\label{Fig:1}	
\end{figure}

We use a 2D phase-field model \cite{moure2021phase,anderson1989colloid,gomez2019review} to study the chemotactic motion of a multicellular system on a planar solid substrate.
Our model describes the chemoattractant dynamics in the extracellular environment, as well as the motion and deformation of each cell caused by chemotaxis and cell-cell interactions.
We denote the problem domain as $\Omega$ and the chemoattractant concentration as $q(\boldsymbol{x},t)$.
We consider a system comprised of $N$ cells and define the location and geometry of the $\alpha$-th cell with the phase field $\phi_\alpha(\boldsymbol{x},t)$, which takes the value $\sim$1 inside the $\alpha$-th cell and $\sim$0 elsewhere (see Fig.~\ref{Fig:1}A).
The evolution of the system is controlled \cite{palmieri2015multiple} by the energy functional $\mathcal{F}[\phi_1,...,\phi_N]=\sum_{\alpha=1}^N\mathcal{F_\alpha} [\phi_1,...,\phi_N]$, where
\begin{eqnarray} 
    \mathcal{F_\alpha}[\phi_1,...,\phi_N] 
    &=& \int_\Omega\gamma\left(\frac{G(\phi_\alpha)}{\epsilon^2} 
    + \abs{\nabla\phi_\alpha}^2 \right)\,{\rm d}\Omega
    +\frac{\lambda}{6A_\alpha}\left( A_{\alpha} - \int_{\Omega}\phi_\alpha^2(3-2\phi_\alpha) {\rm d}\Omega \right)^2 \nonumber\\
    &+& \int_{\Omega}\textit{g}\frac{15}{\epsilon^2}\sum_{\beta \neq \alpha }^N \phi_\alpha^2\phi_\beta^2\,{\rm d}\Omega.
 \label{Eqn:energy-functional}
\end{eqnarray}
The functional $\mathcal{F}_\alpha$ accounts for the membrane surface tension, cell volume (area in 2D) conservation, and cell-cell repulsion, whose strengths are controlled, respectively, by the parameters $\gamma$, $\lambda$, and $g$.
The surface tension contribution makes use of the double-well potential $G(\phi_\alpha)=30\phi_\alpha^2(1-\phi_\alpha)^2$; see \cite{gomez2017computational}.
The parameter $\epsilon$ is proportional to the thickness of the diffuse interface (see Fig.~\ref{Fig:1}A) and ${A}_\alpha$ is the area of cell $\alpha$, which may change over time to capture cell growth, deformation and division. For more details on our modeling approach to cell growth, division, and death see the Supplementary Information.
If we denote the  velocity of the $\alpha$-th cell as $\boldsymbol{u}_\alpha$, we can express the evolution equation for $\phi_\alpha$ and the velocity $\boldsymbol{u}_\alpha$ as
%
%
%
\begin{equation}
    \frac{\partial{\phi_\alpha}}{\partial{t}}
    + \boldsymbol{u}_\alpha\cdot\nabla\phi_\alpha 
    = -\Gamma \frac{\delta \mathcal{F}}{\delta\phi_\alpha}, 
    \quad\quad \quad\quad
    \boldsymbol{u}_\alpha
    =  \boldsymbol{u}_\alpha^A 
    + \frac{60g}{\xi\epsilon^2}\int_{\Omega}\phi_\alpha\nabla\phi_\alpha\sum_{\beta \neq \alpha }^N\phi_\beta^2 \,\text{d}\Omega,
 \label{Eqn:phase-field}
\end{equation} 
where the parameter $\Gamma$ controls the overall kinetics of the driving forces in the functional $\mathcal{F}$, $\xi$ is a constant that represents friction between cells, $\boldsymbol{u}_\alpha^A$ is the active velocity, which we assume uniform in space, and $\frac{\delta \mathcal{F}}{\delta \phi_\alpha}$ is the variational derivative of $\mathcal{F}$ with respect to $\phi_\alpha$; see Methods.
Thus, $\boldsymbol{u}_\alpha$ is uniform in space and has two components:
a passive velocity that emanates from cell-cell interactions ($\xi$-term, see \cite{palmieri2015multiple,wise2008three} for more details) and an active velocity $(\boldsymbol{u}_\alpha^A)$ that describes a persistent random walk biased by the chemoattractant.
In particular, the direction of motion of cell $\alpha$ is biased depending on the characteristic gradient and average concentration of chemoattractant detected by the cell (more details in Methods).
The evolution equation for the extracellular chemoattractant is written as
\begin{equation}
    \frac{\partial (\varphi q)}{\partial t} 
    = \nabla\cdot(D_q\varphi\nabla q)
    + \delta_\varphi b_q
    - \delta_\varphi K_q\frac{q}{K_d+q} 
    - \varphi r_q q ,
     \label{Eqn:chemmoat}
\end{equation}
where the function $\varphi(\boldsymbol{x},t)$, defined as $\varphi = 1 - \sum_{\alpha=1}^N \phi_\alpha$, localizes the extracellular environment and $\delta_\varphi(\boldsymbol{x},t)$, defined as $\delta_\varphi = \frac{30}{\epsilon}\varphi^2(1-\varphi)^2$, localizes the cells membrane (see Fig.~\ref{Fig:1}A).
Eq.~\eqref{Eqn:chemmoat} models chemoattractant diffusion through the extracellular environment, production and consumption on the cells membrane, and degradation, which are controlled by the diffusion coefficient $(D_q)$, and the production $(b_q)$, consumption $(K_q)$, and degradation $(r_q)$ rates, respectively. $K_d$ is the disassociation constant.
Note that the different transport and biochemical processes are localized to the extracellular environment or the cell membrane by including the variable $\varphi$ or $\delta_\varphi$, respectively, into the corresponding terms in Eq.~\eqref{Eqn:chemmoat}. In the limit $\epsilon\to0$, the localization approach is equivalent to a system of PDEs on moving domains with variationally consistent interface conditions \cite{li2009solving}.

\subsection*{\textbf{Dynamic cluster field}}


Conventional multicellular phase-field modeling strategies involve the solution of a separate equation ($\phi_\alpha$) for each cell $\alpha$, which limits the total number of cells $N$ due to high computational costs.
Here, we bypass this limitation by proposing a novel algorithm based on two premises:
(i) We only need to solve Eq.~\eqref{Eqn:phase-field} near cell $\alpha$ because $\partial\phi_\alpha/\partial t = 0$ far from the cell. 
(ii) We can use a single phase field to represent multiple cells that are distant from each other.
This is true as long as we keep track of the area-conservation term and velocity $\boldsymbol{u}_\alpha$ of each cell $\alpha$ in Eq.~\eqref{Eqn:phase-field}.
However, when two cells are close to each other, they must be captured by different phase fields in order to compute cell-cell interaction forces ($g$-terms in Eq.~\eqref{Eqn:phase-field}).\\
%

Premise (i) allows us to solve Eq.~\eqref{Eqn:phase-field} on a \emph{computational window} $\Omega_\alpha\subset \Omega$ defined as the circle of radius $R_\alpha+1.5\epsilon$ centered at cell $\alpha$, where $R_\alpha$ is the equivalent radius of cell $\alpha$.
Although $\Omega_\alpha$ is usually a very small portion of $\Omega$, this definition guarantees that $\Omega_\alpha$ encloses cell $\alpha$ and $\phi_\alpha \approx 0$ outside $\Omega_\alpha$ (see Fig.~\ref{Fig:1}B). 
Premise (ii) allows us to group distant cells into a single \emph{cluster field} $\mathbb{C}_i(\boldsymbol{x},t)$.
Cells within cluster $i$ must maintain a minimum pairwise distance, which we achieve by preventing the overlap of their computational windows.
To define our formulation in terms of cluster fields we rename each phase field $\phi_\alpha$ as $\phi_{i,j}$, where index $i$ identifies clusters and index $j$ identifies cells within each cluster.
If we denote the number of cells in each cluster $i$ as $N_i$, we can define the $i$-th cluster field as $\mathbb{C}_i = \sum_{j=1}^{N_i}\phi_{i,j}$.
Note that $N_i$ and the total number of clusters $(N_c)$ and cells $(N)$ may change due to rearrangements in the spatial distribution of cells, cell division, and cell death, while $\sum_{i=1}^{N_c} N_i = N$ always holds true.
We also rename the computational windows $\Omega_\alpha$ as $\Omega_{i,j}$, in consistentency with the $\phi_{i,j}$ notation, and group these computational windows as $\Omega^{\mathbb{C}}_i=\cup_{j=1}^{N_i}\Omega_{i,j}$.
Thus, we can replace $\phi_\alpha$ by $\mathbb{C}_i$ in Eqs.~\eqref{Eqn:energy-functional} and \eqref{Eqn:phase-field} and solve Eq.~\eqref{Eqn:phase-field} in the computational window $\Omega^{\mathbb{C}}_i$, which reduces the number of PDEs from $N$ to $N_c$.
%
To compute $\boldsymbol{u}_\alpha$ and the area-conservation term in Eq.~\eqref{Eqn:phase-field} for each cell $\alpha$, we track the center of each cell (and, hence, $\Omega_\alpha$), which permits us to use $\mathbb{C}_i$ to compute these terms by restricting the integrals to $\Omega_\alpha$ (more details in Methods).
Finally, the marker $\varphi$ in Eq.~\eqref{Eqn:chemmoat} is rewritten as $\varphi=1-\sum_{i=1}^{N_c}\mathbb{C}_i$.
\\

This cell grouping strategy fails when cells within the same cluster approach each other because cell-cell repulsive forces cannot be computed, which would violate premise (ii). 
%
%
To tackle this issue, we implement a \emph{reallocation strategy} where we reallocate one (or more) of the approaching cells to a different cluster.
%
This procedure involves the cluster exchange between one of the approaching cells and a distant cell from a different cluster, such that the computational windows of cells in the same cluster do not overlap after reallocation (see Fig.~\ref{Fig:1}C).
When cluster exchange is not possible due to the spatial distribution of cells, we create an additional cluster field to capture all pairwise cell-cell interactions and modify $N_c$ accordingly.
%
%
Our reallocation procedure maintains cell position and velocity. 
We refer to the coupled cell grouping and reallocation strategy as \emph{Dynamic Cluster Field} (DCF), which effectively reduces the computational cost of high-resolution simulations of multicellular systems with a large number of cells. The DCF method extends and generalizes the bounding box and reassignment algorithms used in the context of grain growth \cite{krill2002computer,suwa2006phase,vanherpe2007bounding} by redefining the dynamics in terms of the cluster field [see Eq.~\eqref{Eqn:cluster-field}], which is critical for an effective calculation of the individual cell velocities and areas (see Methods). 
For more details about DCF and the numerical implementation, see Methods.

\subsection*{\textbf{Cell co-attraction and proliferation of PC12 cells}}  

PC12 is a cell line that originates from a pheochromocytoma of the rat adrenal medulla. 
These cells are widely used in neurobiological research because they are homogeneous, exhibit rapid growth and differentiation, and secrete dopamine \cite{meldolesi1983effect, wiatrak2020pc12}.
Understanding chemotaxis in PC12 cells is important in neurodegenerative diseases \cite{wiatrak2020pc12}. 
Here, we focus on the experiments in \cite{eyiyurekli2008computational}, which study the aggregation and proliferation of PC12 cells in a Petri dish via self-generated chemoattractant. 
The aggregation process depends on the complex dynamics of the guiding signal in the extracellular space, including how cells push the chemoatractant toward the center of the aggregate as they assemble. Thus, a quantitative understanding of this experimental system requires a detailed description of the chemoattractant dynamics accounting for the time evolution of the extracellular space.
We demonstrate the ability of our model to quantitatively interrogate this non-confluent multicellular system. 
%
We perform 2D simulations on a periodic rectangle of size $720 \times 550$ \SI{}{\micro\meter^2}.
We initially place 39 circular cells grouped in 13 cluster fields.
All cells have an initial radius of \SI{6}{\micro\meter} and are placed at random locations, forming a few cell aggregates.
The cells are placed such that the number of cells per unit area and the aggregate size distribution match the experiment.
We also assume that, at the beginning of the simulation, all cells have the same physical and biochemical properties but different ages.
The initial cell ages are sampled from a uniform random distribution in the interval $[0,24\text{ hours}]$.
We run 4 simulations with different samples of the initial age and initial location of the cells.
In our simulations, the extracellular environment does not contain any chemoattractant at the initial time.
After the simulation starts, the cells secrete chemoattractant through their membranes. 
The internalization of chemoattractant by the cells is neglected. %
Our model considers cell growth, division, and death using information specific to PC12 cells as reported in \cite{eyiyurekli2008computational}.
After cell division, the new cells are in an early-growth stage for $\SI{6}{\hour}$ and cannot divide. After the early-growth stage, cells may undergo division until they reach an age of $\SI{18}{\hour}$. If no division has occurred at this point, cells enter the apoptosis stage, in which they neither grow nor divide. Cells die at age $\SI{24}{\hour}$.
During the early-growth and apoptosis stages, the cells gradually acquire and lose their ability to sense and secrete chemoattractant.
For a detailed description of the cell growth, division, death, and chemotaxis modules of the algorithm, see the Supplementary Information.
\\

%


%
%
%
%

\begin{figure}[h]
	\centering
	\includegraphics[width=0.95\linewidth]{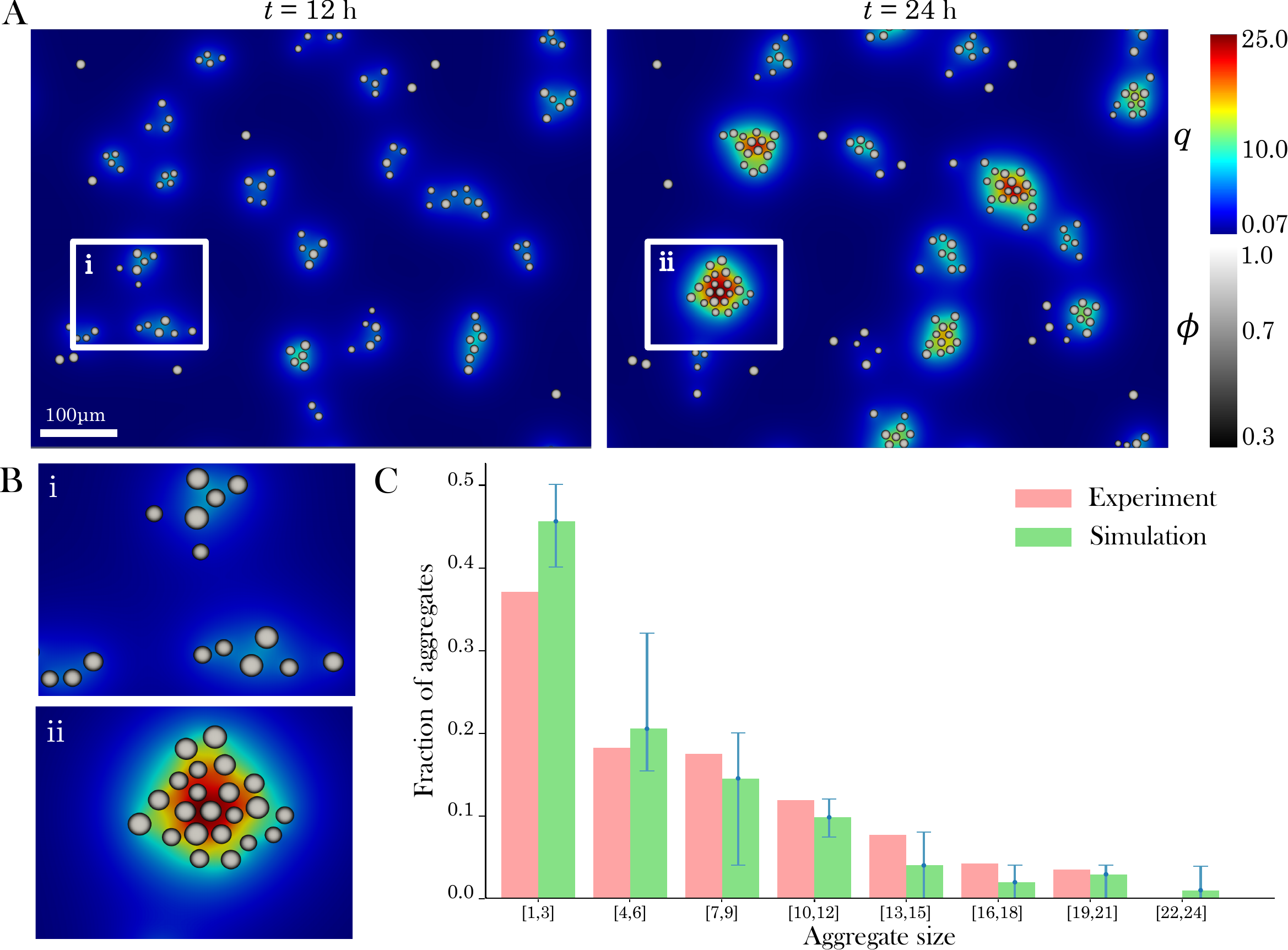}
	\caption{ {\small \textbf{ Cell aggregation driven by autologous chemotaxis.} 
    \textbf{(A)} Cells and chemoattractant distribution at $t=\SI{12}{\hour}$ and $t=\SI{24}{\hour}$.  The unit of $q$ in the plot is \SI{}{\nano\Molar}.
    \textbf{(B)} Zoom-in showing the merging of three aggregates in inset i ($t=\SI{12}{\hour}$) to form one big aggregate in inset ii ($t=\SI{24}{\hour}$).
    We display phase-field ranging from 0.3 to 1 to effectively visualize the chemoattractant distribution in the extracellular region, where $\varphi$ spans from 0.7 to 1.
    \textbf{(C)} Aggregate size distribution at $t=\SI{24}{\hour}$. 
    Experiment observations in red and simulation results in green.
    The aggregate size is divided in 3-cell intervals.
    The error bar indicates the minimum and maximum values observed in our simulations. 
    We set $D_q = \SI{70}{\micro\meter^2\,\second^{-1}}$, $b_q=\SI{18.8}{\nano\Molar\,\micro\meter\,\second^{-1}}$, $r_q=\SI{0.108}{\second^{-1}}$, $\overline{u}_A=\SI{0.04167}{\micro\meter\,\second^{-1}}$, and $\overline{\Delta T}=\SI{60}{\second}$.
    }}
	\label{Fig:coattraction}	
\end{figure}
Figure~\ref{Fig:coattraction}A shows the cells distribution and chemoattractant concentration at $t=\SI{12}{\hour}$ and $\SI{24}{\hour}$ for one simulation. 
The results indicate that aggregate formation and growth are caused by 3 main mechanisms:
(1) Higher chemotactic gradients generated by cell aggregates attract isolated cells.
(2) Merging of different aggregates driven by chemotaxis (the 3 small aggregates in Fig.~\ref{Fig:coattraction}B, inset i form the larger
aggregate in Fig.~\ref{Fig:coattraction}B, inset ii). 
(3) Division of cells within an aggregate. 
As aggregates form, peaks of chemoattractant concentration emerge at their centers (see Fig.~\ref{Fig:coattraction}A, inset ii).
The chemoattractant concentration at the center of the aggregate increases as the aggregate size grows because more cells release chemoattractant in a small region and the cells within the aggregate slow down chemoattractant diffusion through the extracellular environment.
This phenomenon leads to higher chemoattractant gradients pointing towards the aggregate center, which hinders the detachment of cells from the aggregate.
Large aggregates exhibit minimal movement because of the same phenomenon;  while small aggregates and isolated cells are highly motile until they merge with a larger aggregate.
We suspect that the absence of adhesion combined with the reduced ability of cells to sense chemotactic gradient during quiescent period following division might have contributed to the gaps between the cells in the aggregate as observed in Fig.~\ref{Fig:coattraction}(A).
Most cells that are isolated at $t=\SI{24}{\hour}$ (see Fig.~\ref{Fig:coattraction}A) have previously detached from a cluster during the early-growth stage.
During the early-growth stage, cell motion transitions from a persistent random walk to chemotactic migration, as cells develop the ability to sense chemical gradients. Thus, newly born cells can move away from the aggregates even in the presence of opposing chemotactic signals. The possibility of resolving chemoattractant dynamics in the time-evolving extracellular space is a unique feature of our algorithm that enables understanding of the complex interplay between the cell location and the attractant transport dynamics that occur in closely-packed cell aggregates.
%
%
%
\\

Fig.~\ref{Fig:coattraction}C shows the aggregate size distribution at $t=\SI{24}{\hour}$ from the experiments (red) and simulations (green), where we also indicated the minimum and maximum values in our simulations with a blue line.
The simulation results are in good agreement with the experimental data \cite{eyiyurekli2008computational} despite our simulations displaying slightly larger aggregates than the experiment.
We believe this mismatch can be reduced by improving the stochastic model for cell division (more details in Supplementary Information).
%
We also observe that the total number of living cells $N$ doubles by $t\sim\SI{12}{\hour}$ and quadruples by $t\sim\SI{24}{\hour}$, suggesting a 12-hour cell division cycle in agreement with experimental observations.
Our results show that our model effectively captures aggregate formation through the combined processes of proliferation and chemotaxis of PC12 cells.

%
%

%
%
%
%

%
%
%
%

\subsection*{\textbf{Multicellular migration through self-generated chemoattractant gradients} } 


Another situation in which cells play an important role in reshaping the distribution of chemoattractants is when they create chemical gradients by breaking down the extracellular chemoattractant \cite{muinonen2014melanoma,sucgang1997null,dona2013directional}.
This mechanism enables cells to collectively migrate over long distances in a robust manner \cite{tweedy2016self,tweedy2020seeing,tweedy2020self}.
Several experiments \cite{tweedy2016self-steer, tweedy2016self,susanto2017lpp3} analyze the migration of non-confluent multicellular systems driven by self-generated chemotactic gradients.
In particular, experiments in \cite{tweedy2016self} use Dictyostelium discoideum (DD), a type of cell widely used to study chemotaxis due to their strong sensing and motile abilities ---DD's can detect gradients of $1$\% difference in chemoattractant concentration across their membrane \cite{zigmond1974mechanisms}.
\\

\begin{figure}[h!]
	\centering
	\includegraphics[width=0.9\linewidth]{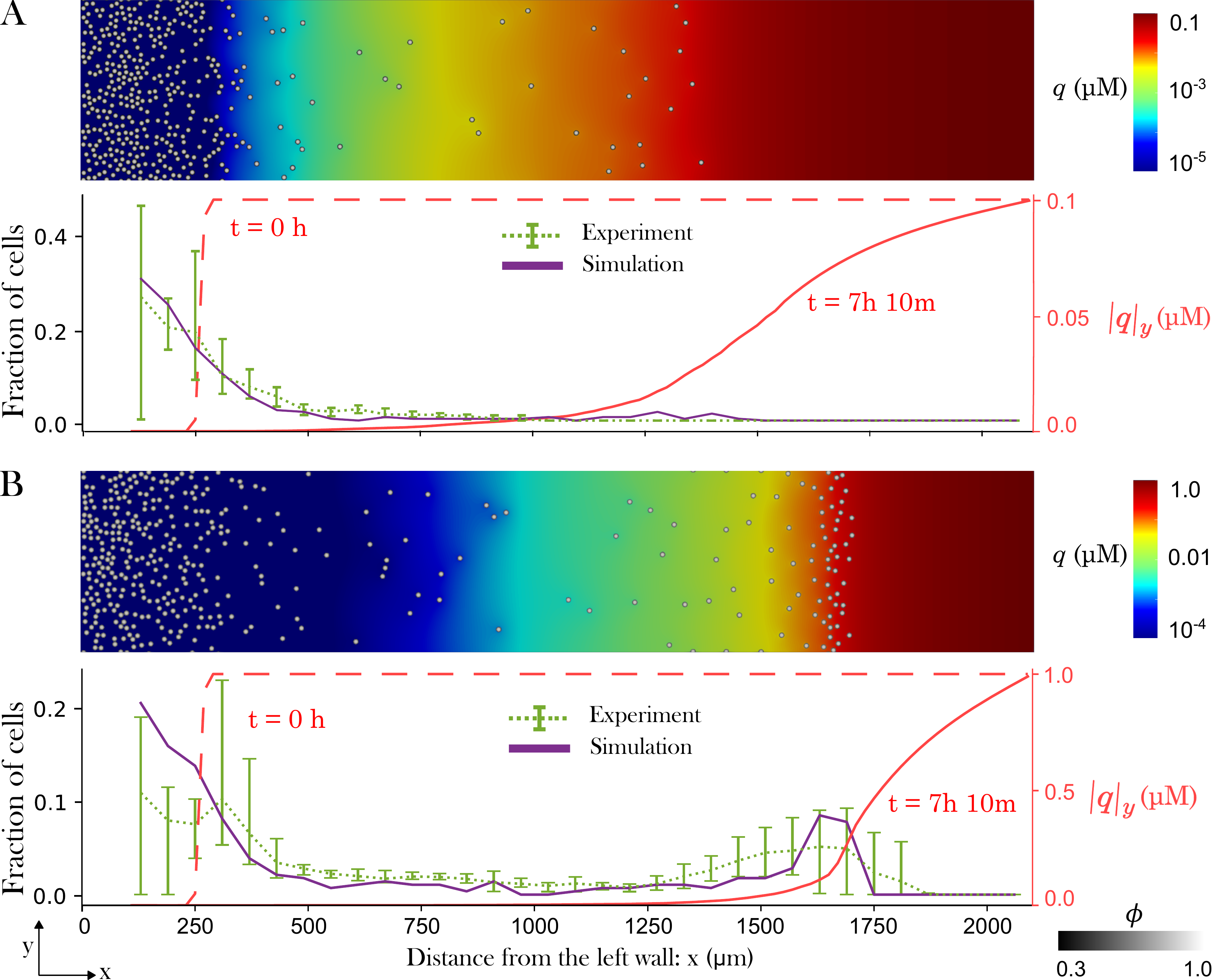}
	\caption{ {\small \textbf{Self-generated chemotaxis of multicellular systems.}
    Evolution of 400 cells in an initial chemoattractant concentration $(q_0)$ of \textbf{(A)} $\SI{0.1}{\micro\Molar}$ and \textbf{(B)} $\SI{1}{\micro\Molar}$. 
    %
    Cells initially occupy the left-hand 250$\SI{}{\micro\meter}$ of the domain and we assume a null $q$ concentration in that region.
    The top half of each panel shows the cell distribution and chemoattractant concentration at $t=\SI{7.166}{\hour}$.
    The bottom half of each panel shows the $y$-averaged chemoattractant concentration $(|q|_y)$ at $t=\SI{0}{\hour}$ (red dashed line) and $t=\SI{7.166}{\hour}$ (red solid line) in our simulations, as well as the fraction of cells along the $x$-axis in our simulations (purple solid line) and the experiment \cite{tweedy2016self} (green dotted line) at the final time.
    We ignore the contribution of cells present between the left boundary and $x = \SI{100}{\micro\meter}$ while computing the fraction of cells to incorporate a well-like region.
    Green bars indicate the minimum and maximum cell fraction observed in the experiment.
    We set $D_q = \SI{12.5}{\micro\meter^2\,\second^{-1}}$, $K_q = \SI{16.5}{\nano\Molar\,\micro\meter\,\second^{-1}}$, $K_d = \SI{50}{\nano\Molar}$,  $\overline{u}_A=\SI{0.072}{\micro\meter\,\second^{-1}}$, and $\overline{\Delta T}=\SI{20}{\second}$.
    }}
	\label{Fig:self_generated}	
\end{figure}

Previous attempts to simulate these experiments have resorted to agent-based models, in which cells have a fixed geometry and are permeable to the chemoattractant. The assumption that cells are permeable to the attractant greatly simplifies the simulation because the computational domain for chemoattractant dynamics has a simple geometry that is fixed in time but misses key elements of the cell-attractant interplay.
Here, we use our model to study the experiments in \cite{tweedy2016self} and analyze the self-driven chemotaxis of DD's under different chemoattractant levels.
The duration of the experiments in \cite{tweedy2016self} is approximately $\SI{7}{\hour}$.
This short time permits us to neglect cell growth, division, and death. 
We also neglect chemoattractant decay and production ($r_q=0$ and $b_q=0$ in Eq.~\eqref{Eqn:chemmoat}) and consider consumption only.
We study a rectangular planar surface of $2100 \times \SI{400}{\micro\meter^2}$ and we initially place 400 cells (grouped in 10 cluster fields) of radius $\SI{5}{\micro\meter}$ at random locations in the 250-$\SI{}{\micro\meter}$ strip in contact with the left boundary.
We assume that the initial chemoattractant concentration is zero in that 250-$\SI{}{\micro\meter}$ strip and is constant in the rest of the domain with concentration $q_0$ (see red-dashed lines in Fig.~\ref{Fig:self_generated}).
We do not monitor the activity of cells within the initial 100-$\SI{}{\micro\meter}$ length from the left boundary to create a well-like area as described in \cite{tweedy2016self}.
We consider periodic boundary conditions in the $y$-direction and fixed $q$ and $\mathbb{C}_i$ on the left and right walls.
We scale the parameter values to perform the simulation on a smaller domain than the experiment (see Supplementary Information). \\

To analyze the influence of the chemoattractant level, we run two simulations with different $q_0$. In particular, we consider $q_0=\SI{0.1}{\micro\Molar}$ and $q_0=\SI{1}{\micro\Molar}$, and we plot the results in Figs.~\ref{Fig:self_generated}A and \ref{Fig:self_generated}B, respectively.
The two panels in Fig.~\ref{Fig:self_generated} show the 2D distribution of cells and $q$ in the top half and the $y$-averaged chemoattractant concentration (red line) and cell fraction (purple line) in the bottom half at $t=\SI{7.166}{\hour}$. The plots also display the range of experimental cell fraction (green-dashed line) observed along the $x$-axis in the bottom half at the end of the experiment.
\\


For $q_0=\SI{0.1}{\micro\Molar}$, our simulation shows a few cells moving to the right wall (Fig.~\ref{Fig:self_generated}A), as observed in the experiment.
These moving (or leading) cells deplete the chemoattractant, which monotonically increases from left to right, with $q\approx 0$ near the left wall (see red line in Fig.~\ref{Fig:self_generated}A).
Thus, cells in that region do not perceive any chemotactic signal and slowly spread out driven by a persistent random walk. 
The magnitude of the chemotactic gradient sensed by the leading cells is low, but high enough to sustain the motion of the leading cells along the channel (see Fig.~\ref{Fig:chem_bias} in Methods).
\\

When the chemoattractant level increases to $q_0=\SI{1}{\micro\Molar}$ (see Fig.~\ref{Fig:self_generated}B), we observe a significantly larger number of cells moving to the right (compare Figs.~\ref{Fig:self_generated}A and B).
The higher chemoattractant concentration enables the presence of larger chemoattractant gradients along longer distances, which results in a more populated wave of leading cells.
In this case, the gradients and average levels of chemoattractant sensed by the leading wave are in an optimal range for chemotaxis (see Fig.~\ref{Fig:chem_bias} in Methods). For this reason, the leading wave exhibits robust migration towards the right wall.
The chemoattractant concentration behind the leading wave is very low so that cells near the left wall perform a non-chemotactic persistent random walk and slowly spread out (Fig.~\ref{Fig:self_generated}B). 
\\

%
%
%

The cell fraction distributions at the final time for both values of $q_0$ show good agreement with the experiment (compare solid-purple and dashed-green lines in Fig.~\ref{Fig:self_generated}), which confirms the potential of our model to capture collective cell migration of non-confluent systems through self-generated chemoattractant gradients.

\subsection*{\textbf{Cell migration in complex chemoattractant environments} } 

Another common scenario in which chemoattractant dynamics are complex and need to be accurately predicted to understand chemotaxis is when there are multiple extracellular species (e.g., an attractant and an enzyme) that interact with each other, giving rise to complex patterns. These patterns emerge from nonlinear dynamics and are sensitive to small perturbations. A cellular system that exhibits this behavior is a conglomerate of vascular mesenchymal cells (VMC), which produce bone morphogenetic protein 2 (BMP-2) and matrix carboxyglutamic acid protein (MGP) \cite{gierer1972theory,garfinkel2004pattern}.
BMP-2 is a molecule recognized for its chemotactic effect on VMC \cite{willette1999bmp}, while MGP is an enzyme that inhibits the production of BMP-2 
\cite{zebboudj2002matrix,bostrom2001matrix}.
The different sizes and, therefore, diffusivity of BMP-2 and MGP lead to the formation of stable Turing patterns \cite{am1952chemical}, where BMP-2 exhibits spatial localization.
Here, we show that accurately predicting pattern formation and chemotaxis requires solving the chemoattractant dynamics in the actual time-evolving extracellular space. In contrast, most models of non-confluent multicellular systems have assumed that the cells are permeable to extracellular components to reduce model complexity by avoiding solving the chemoattractant dynamics on a moving domain.
Here, we quantify the impact of this assumption by studying cell pattern formation in two scenarios: (A) The cells are assumed to be permeable to the extracellular components; and (B) The dynamics of extracellular species occur in the time-evolving extracellular environment. These two scenarios can be studied in our computational model by using the extracellular marker $\varphi$ (see Eq.~\eqref{Eqn:chemmoat}). In case A, we take $\varphi=\varphi_\text{A}=1$, while in case B we use $\varphi=\varphi_\text{B}=1-\sum_{i=1}^{N_c}\mathbb{C}_i$. 
The interactions between BMP-2 (the chemoattractant or activator) and MGP (the inhibitor) can be modeled with the reaction-diffusion equations \cite{gierer1972theory,garfinkel2004pattern,schnorr2023learning}:
\begin{equation}
    \frac{\partial (\varphi q_a)}{\partial t}
    = \nabla\cdot(D_{a}\varphi\nabla q_a) 
    + \varphi \eta \bigg(\frac{q_a^2}{q_i(1+k q_a^2)} 
    - r_a q_a + S \bigg),
    \label{Eqn:activator}
\end{equation}
\begin{equation}
    \frac{\partial (\varphi q_i)}{\partial t} 
    = \nabla\cdot(D_{i}\varphi\nabla q_i)
    + \varphi \eta \big( \chi q_a^2 
    - r_i q_i \big),
    \label{Eqn:inhibitor}
\end{equation}
where $q_j$, $D_j$ and $r_j$, for $j=\{a,i\}$, are the concentration, diffusion coefficient, and decay rate, respectively, of the activator $(a)$ and inhibitor $(i)$.
The parameter $\eta$ controls the overall kinetic rate, $k$ governs the negative feedback between $q_a$ and $q_i$, $\chi$ controls the inhibitor production, and $S$ is the activator source.
Because $D_i>D_a$ and $q_i$ decreases the production of $q_a$ (see $k$-term in Eq.~\eqref{Eqn:activator}) the system of equations leads to the formation of patterns in the activator concentration.
Eqs.~\eqref{Eqn:activator} and \eqref{Eqn:inhibitor} assume spatially uniform production of $q_a$ and $q_i$ in the region defined by $\varphi$, rather than localizing the production to the cells membrane (as done in \cite{garfinkel2004pattern,painter1999stripe}).
This simplification allows us to study the influence of cell permeability in a simpler scenario.\\
%
%

\begin{figure}[h!]
	\centering
	\includegraphics[width=\linewidth]{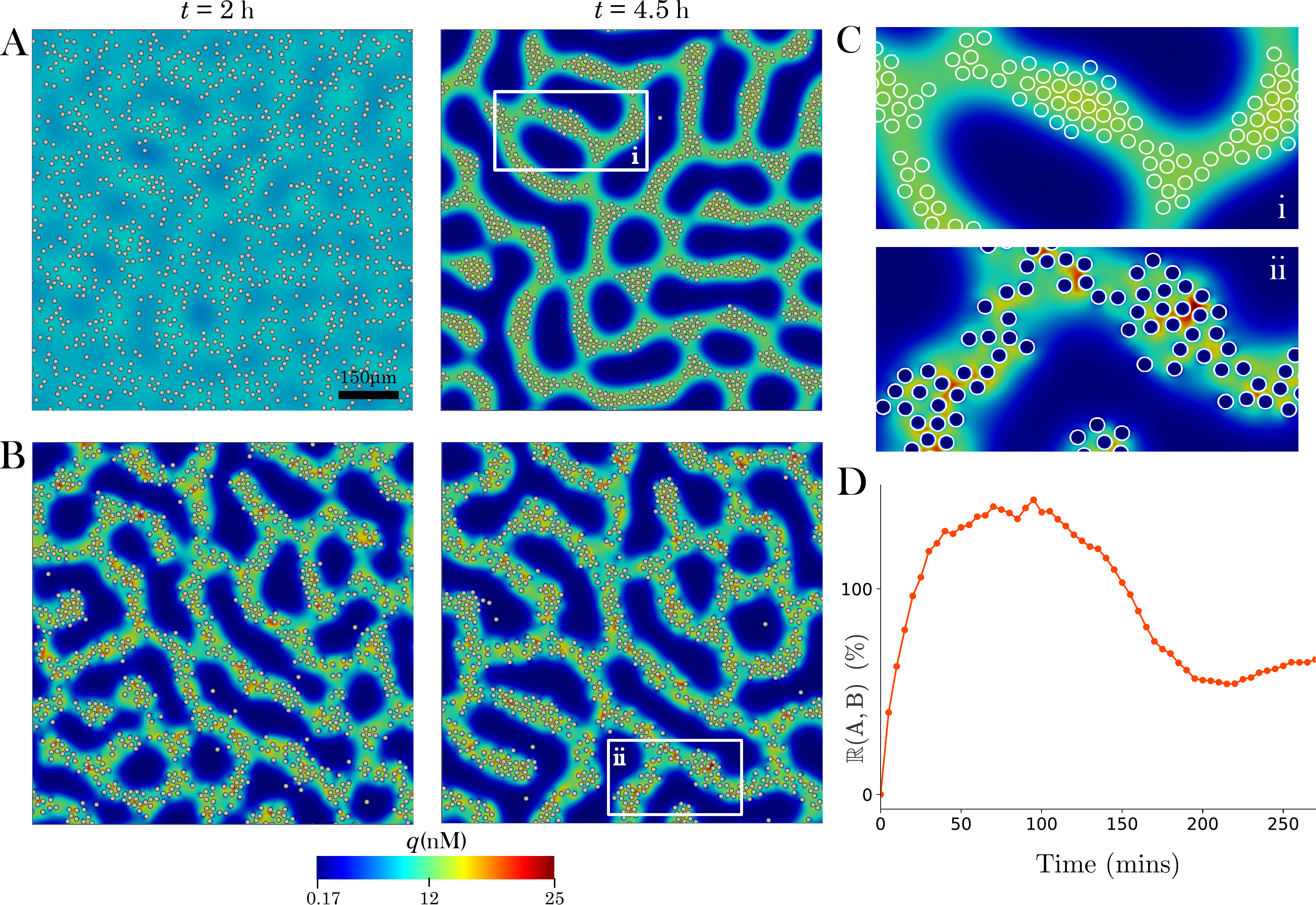}
	\caption{ {\small \textbf{Impact of cell permeability to extracellular components during Turing pattern formation.} 
    Cells and activator $q_a$ distribution for \textbf{(A)} case A (cells permeable to extracellular compounds) and \textbf{(B)} case B (cells not permeable) at $t=\SI{2}{\hour}$ (left) and $t=\SI{4.5}{\hour}$ (right).
    \textbf{(C)} Zoom-in of regions of interest in case A (i, top) and case B (ii, bottom). 
    White lines represent the cell membrane. The colormap shows the $q_a$ distribution outside and inside the cells.
    \textbf{(D)} Time evolution of $\mathbb{R}_{\text{A},\text{B}}$ (defined in the main text).
    We set $D_i = \SI{30}{\micro\meter^2\,\second^{-1}}$, $D_a = \SI{0.3}{\micro\meter^2\,\second^{-1}}$, $r_a = r_i = 1$, $k =\SI{0.72}{\nano\Molar^{-2}}$, $\chi=\SI{1}{\nano\Molar^{-1}}$, $S=\SI{0.02}{\nano\Molar}$, and $\eta = \SI{0.00469}{\second^{-1}}$, $\overline{u}_A=\SI{0.04167}{\micro\meter\,\second^{-1}}$, and $\overline{\Delta T}=\SI{20}{\second}$ 
    }}
    \vspace{-0.1in}
    \label{Fig:Turing_pattern}	
\end{figure}

In this instantiation of our model, Eqs.~\eqref{Eqn:activator} and \eqref{Eqn:inhibitor} replace Eq.~\eqref{Eqn:chemmoat}. The equations governing cell dynamics [Eqs.~\eqref{Eqn:energy-functional} and \eqref{Eqn:phase-field}] remain identical, but chemotaxis is directed by $q_a$ instead of $q$.
We study the evolution of a multicellular system for $\SI{4.5}{\hour}$, which permits us to neglect cell growth, division, and death.
For both cases A and B, we consider a computational domain of size $950\times\,\SI{950}{\micro\meter^2}$ with periodic boundary conditions.
We initially place 1040 VMC grouped in 13 clusters. 
All cells have an initial radius of $\SI{6}{\micro\meter}$ and a random initial location. 
The variables $q_a$ and $q_i$ are zero in the intracellular space at the initial time.
The initial conditions for $q_a$ and $q_i$ in the extracellular space are given by $q_a = \overline{q}_a + r_a$ and $q_i = \overline{q}_i + r_i$. Here, $\overline{q}_a = \SI{8.36}{\nano\Molar}$, $\overline{q}_i = \SI{7.81}{\nano\Molar}$ and $r_a$, $r_i$ represent random perturbations sampled from a uniform distribution such that $q_a$ and $q_i$ deviate up to 10\% from their average values $\overline{q}_a$ and $\overline{q}_i$.
\\

Figures~\ref{Fig:Turing_pattern}A and \ref{Fig:Turing_pattern}B show the pattern formation process for cases A and B, respectively.
The results at $t=\SI{2}{\hour}$ for case A, which assumes that the chemoattractant can diffuse freely into the cells interior, do not show any discernable pattern (Fig.~\ref{Fig:Turing_pattern}A-left).
However, in case B the model prediction at $t=\SI{2}{\hour}$ already shows a strong pattern for cells and activator (Fig.~\ref{Fig:Turing_pattern}B-left).
The faster pattern formation in case B is caused by the motion of the cells. The combined action of cell motion and activator pattern formation leads to a positive feedback effect that accelerates the aggregation. Cells displace the chemoattractant, which accumulates in the cell's front, and concentrate the chemoattractant in certain regions, leading to a stronger chemical signal. This phenomenon is not observed when cells are permeable to extracellular components (case A). 
At the final time $(t=\SI{4.5}{\hour})$, both cases display distinctive patterns (Figs.~\ref{Fig:Turing_pattern}A-right and B-right). 
However, in case A, the cells have a more uniform spatial distribution within the activator pattern. One manifestation of this is that the pairwise distance of cells within the pattern is more uniform in case A than in case B. We hypothesize that this behavior is caused by the higher complexity of the chemoattractant distribution in case B. The cells trap chemoattractant in small regions, leading to a larger number of local maxima in the chemical signal, which produces a more random motion within the activator pattern; compare insets plotted in Fig.~\ref{Fig:Turing_pattern}C. \\
%
%

%


A more quantitative comparison between the two cases can be made by computing
\begin{equation}
    \mathbb{R}_{\text{A},\text{B}}(t)=\frac{ \sum_{\alpha=1}^N \left[\big| \textbf{c}_\alpha^B(t) - \textbf{c}_\alpha^B(0) \big|^2 - \big| \boldsymbol{c}_\alpha^A(t) - \boldsymbol{c}_\alpha^A(0) \big|^2 \right]}{\sum_{\alpha=1}^N \big| \boldsymbol{c}_\alpha^A(t) - \boldsymbol{c}_\alpha^A(0) \big|^2 }. 
    \label{Eqn:Rab}
\end{equation}
Here, $\boldsymbol{c}_\alpha^A(t)$ is the center of cell $\alpha$ at time $t$ for case A and $\boldsymbol{c}_\alpha^B(t)$ is the same quantity for case B.
Thus, $\mathbb{R}_{\text{A},\text{B}}$ represents the relative difference of the cell's mean square displacement between cases A and B.
Positive values of $\mathbb{R}_{\text{A},\text{B}}$ imply larger cell displacements with respect to the initial position in case B, compared to case A.
%
%
We plot the time evolution of $\mathbb{R}_{\text{A},\text{B}}$ in Fig.~\ref{Fig:Turing_pattern}D.
The plot shows that $\mathbb{R}_{\text{A},\text{B}}$ increases rapidly at first, reaching a maximum at $t\approx\SI{2}{\hour}$. 
This increase is compatible with the faster pattern formation observed in case B: While the cells in case A still sense an approximately uniform activator distribution at $t\approx\SI{2}{\hour}$ that limits their mobility, the cells in case B have undergone significant displacement and formed a well-defined pattern.
In the time interval from $t\approx\SI{2}{\hour}$ to $t\approx\SI{200}{\minute}$, the cells in case A form a pattern, whereas the cells in case B remain within the previously established activator pattern. This explains why $\mathbb{R}_{\text{A},\text{B}}$ decreases in this time interval; see Fig.~\ref{Fig:Turing_pattern}D. 
For both cases, after $t\approx\SI{200}{\minute}$ the cells display similarly low levels of motion within the activator pattern and therefore $\mathbb{R}_{\text{A},\text{B}}$ remains roughly constant. \\

%

%

Our results show that assuming that the chemoattractant can diffuse into the intracellular space leads to an important underestimation of the ability of the cells to form patterns, especially at early times. 
This emphasizes the importance of solving the chemoattractant dynamics in the extracellular space and highlights the advantages of the DCF approach to model the chemotaxis of nonconfluent multicellular systems.



\section*{Discussion}

Despite significant recent advances, our understanding of collective chemotaxis of eukaryotic cells remains limited.
Due to the complex interplay between the migrating cells and the attractant distribution, it is difficult to control or measure precisely the chemoattractant distribution, which hinders our mechanistic understanding.
In this paper, we proposed the DCF method, a new algorithm that provides a high-fidelity resolution of the attractant dynamics accounting for the time-evolving geometry and topology of the extracellular space, chemoattractant production and uptake by the cell as well as cell-cell interactions and cell deformation.
\\

We illustrated the potential of our approach by performing unprecedented high-fidelity simulations with $N=\mathcal{O}(1000)$ migrating cells with chemoattractant diffusing and reacting in the extracellular environment.
Our results show that the transport dynamics of the attractant are highly influenced by the changes in extracellular space produced by cell migration, especially when cells are highly packed.
In this situation, the effective space for chemoattractant diffusion is greatly reduced leading to larger gradients that would not occur otherwise.
\\

Our algorithm enabled computation with a $60$-fold increase in the number of cells compared to the existing methods that resolve chemoattractant and cell dynamics with the same level of fidelity.
We tested our computational model by comparing the results with two experiments.
The agreement in the results from experiments and simulations showcases the predictive capabilities of the computational model under a variety of scenarios including co-attraction of cells and migration through self-generated chemoattractant gradients. 
Although approaches such as the Agent-based method, Cellular Potts Method, and Spheroidal model can simulate collective cell migration of a very large number of cells at lower computational cost than DCF, they provide lower resolution of the chemoattractant dynamics because they assume the chemoattractant can freely permeate the cell.
Because the cell's trajectories are sensitive to small chemical gradients lack of accuracy in the chemoattractant distribution can compromise the predictions of the cell's motion.
\\

Our model can help address multiple open questions that involve collective chemotaxis in the central nervous system \cite{theveneau2010collective}, immune response \cite{malet2015collective}, and cancer metastasis \cite{roussos2011chemotaxis}.
The proposed computational model can be extended to incorporate cell-cell adhesion, contact inhibition of locomotion, and various other relevant factors to further enhance our understanding of collective cell migration dynamics.

\section*{Methods}

\subsection*{Chemotactic cell velocity}

\begin{figure}[h!]
\centering
\includegraphics[width=0.9\linewidth]{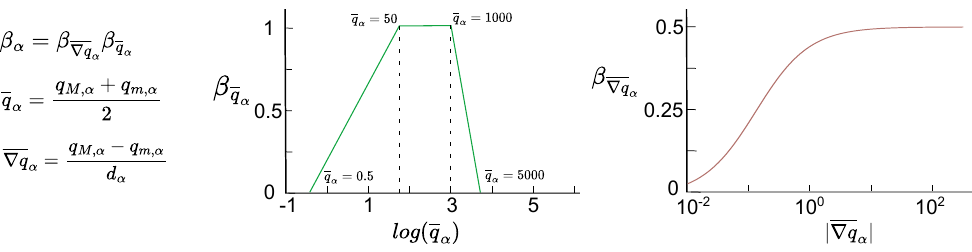}
\caption{ {\small 
The characteristic gradient $(\overline{\nabla q}_\alpha)$ and average level $(\overline{q}_\alpha)$ of chemoattractant are defined as a function of the maximum $(q_{M,\alpha})$ and minimum $(q_{m,\alpha})$ chemoattractant concentration on the cell membrane and the distance $d_\alpha$ between the locations of $q_{M,\alpha}$ and $q_{m,\alpha}$. The chemotactic bias $\beta_\alpha$ that is applied to the cell velocity orientation depends on $\overline{q}_\alpha$ and $\overline{\nabla q}_\alpha$ through the functions $\beta_{\overline{q}_\alpha}$ (left) and $\beta_{\overline{\nabla q}_\alpha}$ (right), respectively, such that $\beta_\alpha=\beta_{\overline{\nabla q}_\alpha}\beta_{\overline{q}_\alpha}$ (see \cite{moure2018three,song2006dictyostelium,fuller2010external,van2010stochastic} for more details). The units of $\overline{q}_\alpha$ and $\overline{\nabla q}_\alpha$ in the plot are $\SI{}{\nano\Molar}$ and $\SI{}{\nano\Molar\,\micro\meter^{-1}}$, respectively.
}}
\label{Fig:chem_bias}	
\end{figure}

The active velocity $\boldsymbol{u}^{A}_\alpha$ in Eq.~\eqref{Eqn:phase-field} is uniform in space and defines a persistent random walk biased by the average chemotactic gradient $(\overline{\nabla q}_\alpha)$ and the average chemoattractant concentration $(\overline{q}_\alpha)$ detected by cell $\alpha$; see definitions in Fig.~\ref{Fig:chem_bias}.
We define $\boldsymbol{u}^{A}_\alpha$ following the stochastic approach described in \cite{moure2021phase}. 
In this approach, $\boldsymbol{u}^{A}_\alpha$ remains constant for a time interval $\Delta T_\alpha$.
When the current interval ends, we compute a new time interval length, as well as a new norm and direction of $\boldsymbol{u}^{A}_\alpha$ for the next time interval based on probability distributions taken from experiments \cite{bosgraaf2009ordered,song2006dictyostelium,fuller2010external,van2010stochastic}.
More precisely, $\vert\boldsymbol{u}^{A}_\alpha\vert$ and $\Delta T_\alpha$ are sampled from uniform random distributions $U(0.8\overline{u}_A,1.2\overline{u}_A)$ and $U(0.8\overline{\Delta T},1.2\overline{\Delta T})$, respectively, where $\overline{u}_A$ and $\overline{\Delta T}$ are experimentally adjusted constants \cite{eyiyurekli2008computational,tweedy2016self,garfinkel2004pattern,moure2021phase}.
These experiments also provide the change in the direction of $\boldsymbol{u}^{A}_\alpha$  between two consecutive intervals for an unbiased persistent random walk, represented by the angle $\theta_\alpha$.
To model chemotaxis, we bias $\theta_\alpha$ towards the chemoattractant gradient $(\overline{\nabla q}_\alpha)$ by rotating it an angle $\beta_\alpha\theta_\alpha$, where $\beta_\alpha$ is defined as $\beta_\alpha=\beta_{\overline{\nabla q}_\alpha}\beta_{\overline{q}_\alpha}$.
Here, $\beta_{\overline{\nabla q}_\alpha}$ and $\beta_{\overline{q}_\alpha}$ depend on $\overline{\nabla q}_\alpha$ and $\overline{ q}_\alpha$, respectively, as described in Fig.~\ref{Fig:chem_bias}.
For a detailed description of the algorithm used to calculate $\boldsymbol{u}^{A}_\alpha$, see \cite{moure2021phase,moure2018three}.
Our algorithms for cell growth, division, and death are described in the Supplementary Information. \\

%
%
%
%
%
%

\subsection*{Numerical implementation}

The variational derivative $\frac{\delta\mathcal{F}}{\delta \phi_\alpha}$ in Eq.~\eqref{Eqn:phase-field} is expressed as
\begin{eqnarray}
 \label{Eqn:variational}
    \frac{\delta\mathcal{F}}{\delta \phi_\alpha} 
    &=& \gamma\left(\frac{60}{\epsilon^2}\phi_{\alpha}(1-\phi_{\alpha})(1-2\phi_{\alpha}) 
    - 2\nabla^2\phi_{\alpha}\right) \nonumber\\
    &-& 2\lambda\left[\frac{\phi_\alpha(1-\phi_\alpha)}{A_{\alpha}}\left(A_{\alpha} 
    - \int_{\Omega}\phi_\alpha^2(3-2\phi_\alpha) \,\text{d}\Omega\right) \right] 
    + g\frac{60}{\epsilon^2}\sum_{\beta \neq \alpha }^N\phi_\alpha\phi_\beta^2. 
\end{eqnarray} 
To simplify the notation, from here on, we denote the expression inside the square brackets which is multiplied by $\lambda$ in Eq.~\eqref{Eqn:variational} as $\Delta \!V_\alpha(\phi_\alpha)$. \\

The velocity $\boldsymbol{u}_\alpha$ (see Eq.~\eqref{Eqn:phase-field}), which is uniform in space, must be tracked for each individual cell even after grouping the cells into clusters.
The calculation of the active component $\boldsymbol{u}_\alpha^A$ is described in the previous section. The passive component results from computing the $\xi$-term in Eq.~\eqref{Eqn:phase-field}. When using the clustering approach, the $\xi$-term is calculated from $\mathbb{C}_i$ and restricting the integral to $\Omega_\alpha$.
We can apply the same strategy to compute the area-conservation term $\Delta \!V_\alpha(\phi_\alpha)$.
%
%
Consistently with our cluster-based indexation, we identify cell $\alpha$ with cell $(i,j)$ and the computational window $\Omega_\alpha$ with $\Omega_{i,j}$.
Thus, we can write
\begin{equation}
    \boldsymbol{u}_{i,j}
    =  \boldsymbol{u}_{i,j}^A 
    + \frac{60g}{\xi\epsilon^2}\int_{\Omega_{i,j}}\mathbb{C}_i\nabla\mathbb{C}_i\sum_{k \neq i }^{N_c}\mathbb{C}_k^2 \,\text{d}\Omega,
    \quad\quad\quad
    \Delta \! V_{i,j}=\frac{\mathbb{C}_i(1-\mathbb{C}_i)}{A_{i,j}}\left(A_{i,j} 
    - \int_{\Omega_{i,j}}\mathbb{C}_i^2(3-2\mathbb{C}_i) \,\text{d} \Omega\right),
    \label{Eqn:velocity_V_ij} 
\end{equation}
where the integral can be computed on $\Omega_{i,j}$ because we keep track of the center of each cell.
We can group the velocity $\boldsymbol{u}_{i,j}$ and the area-conservation term $ \Delta \! V_{i,j}$ of cells within the same cluster $i$ into piecewise functions $\boldsymbol{u}^{\mathbb{C}}_i$ and $ \Delta \! V^\mathbb{C}_i$, respectively, such that
%
\begin{equation}
    \boldsymbol{u}^\mathbb{C}_i(\boldsymbol{x},t)=\begin{cases}
          \boldsymbol{u}_{i,j}(t) \: & \text{if} \; \boldsymbol{x} \in \Omega_{i,j}, \\
          0 \: & \text{otherwise}, \ 
     \end{cases}
     \quad\quad\quad\quad\quad
    \Delta \! V^{\mathbb{C}}_i(\mathbb{C}_i)=\begin{cases}
           \Delta \! V_{i,j}(\mathbb{C}_i) \: &\text{if} \; \boldsymbol{x} \in \Omega_{i,j}, \\
          0 \: &\text{otherwise}. \ 
     \end{cases}
     \label{Eqn:velocity_V_grouping}
\end{equation}
Note that $\boldsymbol{u}^{\mathbb{C}}_i$ and $\Delta \! V^{\mathbb{C}}_i$ are well-defined because our reallocation strategy prevents overlap of computational windows that belong to the same cluster. 
Thus, we can replace $\phi_\alpha$, $\boldsymbol{u}_\alpha$, and $\Delta \! V_\alpha$ by $\mathbb{C}_i$, $\boldsymbol{u}^{\mathbb{C}}_i$, and $\Delta \! V^{\mathbb{C}}_i$ respectively, in Eq.~\eqref{Eqn:phase-field}, which yields 
\begin{equation}
    \frac{\partial{\mathbb{C}_i}}{\partial{t}} + \boldsymbol{u}^\mathbb{C}_i\cdot\nabla\mathbb{C}_i
    = 
    -\Gamma\left[\gamma\left(\frac{60}{\epsilon^2}\mathbb{C}_i(1-\mathbb{C}_i)(1-2\mathbb{C}_i) 
    - 2\nabla^2\mathbb{C}_i\right)
    -2\lambda \Delta \! V^{\mathbb{C}}_i(\mathbb{C}_i)
    + g\frac{60}{\epsilon^2}\sum_{k \neq i }^{N_c} \mathbb{C}_i \mathbb{C}_k^2 \right].
 \label{Eqn:cluster-field}
\end{equation} 
The numerical solution of the chemoattractant dynamics [Eq.~\eqref{Eqn:chemmoat}] may be unstable because the time derivative of $q$ is multiplied by the extracellular marker $\varphi$, which vanishes inside the cells.
%
%
To bypass this issue, we introduce the variable $\hat{q} = \varphi q$ and rewrite Eq.~\eqref{Eqn:chemmoat} as
\begin{equation}
    \frac{\partial\hat{q}}{\partial t} 
    = \nabla \cdot(D_q\nabla\hat{q}) 
    - \nabla\cdot\left(\frac{D_q\hat{q}\nabla\varphi}{\varphi}\right) 
    + \delta_\varphi b_q
    - \delta_\varphi K_q \frac{\hat{q}}{\varphi K_d +\hat{q}} 
    -  r_q \hat{q}.
    \label{Eqn:chemmoat_hat}
\end{equation}
In our numerical implementation we solve Eq.~\eqref{Eqn:chemmoat_hat} instead of Eq.~\eqref{Eqn:chemmoat}. This increases the robustness of the algorithm and leads to more stable solutions.
%


\paragraph*{Spatial discretization}

Our spatial discretization is based on Isogeometric Analysis which is a generalization of the finite element method (FEM), which utilizes splines as basis functions \cite{cottrell2009isogeometric}.
%
%
%
%
%
%
We obtain the weak form of the problem by multiplying Eqs.~\eqref{Eqn:cluster-field} and \eqref{Eqn:chemmoat_hat} with weight functions, integrating on the computational domain $\Omega$, and integrating by parts considering appropriate boundary conditions.
Let $\boldsymbol{U} = [\{\mathbb{C}_i\}_{i=1}^{N_c}, \hat{q}]^T = [U_1,...,U_{N_c+1}]^T$ be the solution vector containing all the unknowns. 
%
%
We define the trial function space as $\mathcal{V}^{\mathbb{U}} \subset [\mathcal{H}^1(\Omega)]^{N_c+1}$, where $\mathcal{H}^1(\Omega)$ is the Sobolev space of square-integrable functions with square-integrable first derivatives in $\Omega$.
Let the weighting function space $\mathcal{W}^{\mathbb{U}}$ be identical to the trial function space $\mathcal{V}^{\mathbb{U}}$.
The problem can be stated as: Find $\boldsymbol{U} \in \mathcal{V}^{\mathbb{U}}$ such that $\forall \boldsymbol{w} \in \mathcal{W}^{\mathbb{U}}$
\begin{equation}
    \int_{\Omega} w_l\frac{\partial{U}_l}{\partial t} \,\text{d} \Omega
    =
    \mathcal{R}_l(w_l,\boldsymbol{U}),
    \quad \quad \text{for }l = 1,2,..,{N_c+1},
    \label{Eqn:weak_form}
\end{equation}
where the residuals $\mathcal{R}_l$ are defined as
%
\begin{gather}
\begin{split}
   \mathcal{R}_l(w_l,\boldsymbol{U})
    =&  - \int_{\Omega}w_l \boldsymbol{u}_l^\mathbb{C}\cdot\nabla\mathbb{C}_l \,\text{d}\Omega 
    - \int_{\Omega}w_l\Gamma\gamma\frac{60}{\epsilon^2}\mathbb{C}_l(1-\mathbb{C}_l)(1-2\mathbb{C}_l) \,\text{d}\Omega 
    - \int_{\Omega}\Gamma \gamma 2 \nabla w_l\cdot\nabla\mathbb{C}_l \,\text{d}\Omega\\
    & + \int_{\Omega}w_l\Gamma 2 \lambda \,\Delta \! V_l^\mathbb{C}(\mathbb{C}_l) \,\text{d}\Omega 
    - \int_{\Omega} w_l \Gamma g \frac{60}{\epsilon^2}\sum_{m \neq l } ^{N_c} \mathbb{C}_l\mathbb{C}_m^2  \,\text{d}\Omega,\;\;\text{ for } l=1,\dots, N_c.
 \end{split}
 \label{Eqn:res_cluster}
\end{gather} 
\begin{gather}
\begin{split}
    \mathcal{R}_{N_c+1}(w_{N_c+1},\boldsymbol{U})
    =& - \int_{\Omega}D_q\nabla w_{N_c+1}\cdot\nabla \hat{q} \,\text{d}\Omega 
    + \int_{\Omega}D_q \frac{\hat{q}}{\varphi^\star} \nabla w_{N_c+1}\cdot\nabla\varphi \,\text{d}\Omega 
    + \int_{\Omega}w_{N_c+1}\delta_\varphi b_q \,\text{d}\Omega \\
    &- \int_{\Omega}w_{N_c+1} \delta_\varphi K_q  \frac{\hat{q}}{\varphi^\star K_d +\hat{q}} \,\text{d}\Omega
    - \int_{\Omega}w_{N_c+1}r_q \hat{q} \,\text{d}\Omega, 
 \end{split}
 \label{Eqn:res_chem}
\end{gather} 
To make the algorithm more stable, in the second and fourth integral of Eq.~\eqref{Eqn:res_chem}, we replaced $\varphi$ by $\varphi^\star$ to avoid the singularity inside the cells, where $\varphi =0$.
We take $\varphi^\star=\max(0.01,\varphi)$. 
This procedure does not affect the solution because the modified term is approximately zero where $\varphi<0.01$.\\

%

%
To discretize the weak form of the problem, we replace the unknowns ${U}_l(\boldsymbol{x},t)$ and weight functions $w_l(\boldsymbol{x})$ with their discrete approximations ${U}_l^h(\boldsymbol{x},t)=\sum_A^{n_b}{U}_{l,A}(t)N_A(\boldsymbol{x})$ and $w_l^h(\boldsymbol{x})= \sum_A^{n_b} w_{l,A} N_A(\boldsymbol{x})$, respectively.
%
%
Here, $n_b$ is the dimension of the discrete space, the $N_A$'s are quadratic $\mathscr{C}^1$-continuous B-splines and the coefficients ${U}_{l,A}$'s are the control variables.
%
%
\\

\paragraph*{Time stepping algorithm} 
To perform time integration, we divide the simulation time into a sequence of intervals $(t_n,t_{n+1})$ with time step $\Delta t=t_{n+1}-t_n$.
Because we are solving a very large system of PDEs, an implicit time integration scheme would be inefficient. The dynamics of the system is stiff due to the singularity of the equations in the extracellular space. This stiffness imposes severe constraints on the time step to be used. Thus, high-order accuracy is not required and we employ the forward Euler method. 
The time-discrete weak form is 
\begin{equation}
    \int_\Omega w_l^h{U}_l^{h,n+1} \,\text{d} \Omega
    =
    \int_\Omega w_l^h{U}_l^{h,n} \,\text{d} \Omega
    + \Delta t \, \mathcal{R}_l(w_l^h,\boldsymbol{U}^{h,n}),
    \quad \quad \text{for }l = 1,2,..,{N_c}+1,
    \label{Eqn:forward_euler}
\end{equation}
where the superscript $n$ indicates time $t_n$.
If we replace $U_l$ and $w_l$ by their discrete approximations in Eq.~\eqref{Eqn:forward_euler} and use the linearity of the residual with respect to the weight function, we obtain
\begin{equation}
    \sum_A^{n_b}\sum_B^{n_b} w_{l,A}{U}_{l,B}^{n+1} \int_\Omega N_A N_B  \,\text{d} \Omega = \sum_A^{n_b}\sum_B^{n_b} w_{l,A}{U}_{l,B}^n \int_\Omega N_A N_B  \,\text{d} \Omega + \sum_A^{n_b} w_{l,A}\Delta t\, \mathcal{R}_l \left( N_A ,\boldsymbol{U}^{h,n} \right),
    \label{Eqn:euler_discretized}
\end{equation}
for $l = 1,2,..,{N_c}+1$.
%
%
If we define the mass matrix as ${M}_{AB} = \int_\Omega N_A N_B \,\text{d} \Omega$, and we use the arbitrariness of the weight functions we can rewrite Eq.~\eqref{Eqn:euler_discretized} in matrix format as 
\begin{equation}
    {M}_{AB} {U}_{l,B}^{n+1} 
    =
     {M}_{AB} {U}_{l,B}^n 
    + \Delta t\, \mathcal{R}_l \left( N_A,\boldsymbol{U}^{h,n} \right), \quad \quad \text{for }l = 1,2,..,{N_c}+1,
    \label{Eqn:euler_matrix}
\end{equation}
Note that for a fixed $l$, $M_{AB}$ is a matrix of size $(n_b \times n_b)$, and ${U}_{l,B}^n$ and $\mathcal{R}_l$ are vectors of size $n_b$.
To speed up the numerical solution of Eq.~\eqref{Eqn:euler_matrix}, we resort to the row-sum mass matrix lumping method \cite{duczek2019critical,duczek2019mass,gravenkamp2020mass}, in which the mass matrix $M_{A,B}$ is approximated by its corresponding lumped mass matrix $\mathcal{M}_{AB}$.
This method simplifies matrix inversion, speeds up the numerics, and reduces memory use. 
The lumped mass matrix is computed as 
\begin{equation}
    \mathcal{M}_{AB}=\begin{cases}
          \sum_{b=1}^{n_b} {M}_{A,b} \: & \text{if} \; A=B, \\
          0 \: & \text{if} \; A\neq B. \ 
     \end{cases}
     \label{Eqn:lumped_matrix}
\end{equation}
Because $\mathcal{M}_{AB}$ is a diagonal matrix, its inverse can be computed at cost $\mathcal{O}(n_b)$, Eq.~\eqref{Eqn:euler_matrix} can be rewritten as 
\begin{equation}
    {U}_{l,B}^{n+1} = 
    {U}_{l,B}^n 
    + \Delta t \, \mathcal{M}_{AB}^{-1}\mathcal{R}_l(N_A,\boldsymbol{U}^{h,n}) \quad \quad \text{for }l = 1,2,..,{N_c}+1.
    \label{Eqn:euler_lumped_matrix}
\end{equation}

After solving Eq.~\eqref{Eqn:euler_lumped_matrix}, we perform the following tasks at $t=t_{n+1}$:

\begin{enumerate}[label={(\arabic*)}]

\item For each cell $\alpha$, compute its area $A_\alpha$, its center $\boldsymbol{c}_\alpha$, its equivalent radius $R_\alpha$, and update its computational window $\Omega_\alpha$.

\item For each cell $\alpha$, if $t^{n+1}$ falls beyond the cell's current time interval $\Delta T_\alpha$, compute $\overline{\grad q}_\alpha$, $\overline{q}_\alpha$, $\beta_\alpha$ and update $\boldsymbol{u}^A_\alpha$ and $\Delta T_\alpha$ for the next interval.

\item If cell division and death are considered: 

\begin{enumerate}[label={(\roman*)}]
\item Update the values of $b_{q}$ and $\beta_\alpha$ if cell $\alpha$ is the early-growth stage or the apoptosis stage.
\item For each cell $\alpha$ out of the early-growth and apoptosis stages, evaluate the division criterion and proceed to divide cell if needed.
\end{enumerate}

\item Cell reallocation: if computational windows of cells in the same cluster overlap, we proceed with the reallocation algorithm described in the main text. The cell-dependent variables are also reallocated.

\item Construct $\boldsymbol{u}^{\mathbb{C}}_{i}$ and $\Delta \! V^{\mathbb{C}}_i$ as described in Eq.~\eqref{Eqn:velocity_V_grouping}.
    
\end{enumerate}

At this point, the algorithm advances to the next time step, i.e., proceeds to solve Eq.~\eqref{Eqn:euler_lumped_matrix} to compute the solution at $t=t_{n+2}$.
Task (3) is explained in the Supplementary Information and only considered in the example shown in Fig.~\ref{Fig:coattraction}.\\

We build our code on top of the scientific libraries PETSc \cite{balay2020petsc} and PetIGA \cite{dalcin2016petiga}. PetIGA adds to PETSc capabilities for NURBS and B-spline analysis as well as integration of forms.

\subsection*{Additional information}

We consider the same element size for all the simulations in the paper. In particular, we construct all meshes with square elements of size $1\times\SI{1}{\micro\meter^2}$. 
The parameter values common for all examples are listed in Table~\ref{tab:parameter_values}.
The rest of the parameters values are listed in the corresponding section in the main text.
Further details of the simulations shown in this work may be found in Supplementary Information.
\begin{table}[H]
    \centering
    \begin{tabular}{|l l c |} 
    \hline
    Symbol & Description & Value  \\ [0.5ex] 
    \hline
    $\epsilon$ & Phase-field interface thickness & $\SI{4}{\micro\meter}$  \\ 
    $\gamma$ & Surface tension coefficient & $\SI{20}{\pico\newton}$ \\
    $ \lambda$ & Area-conservation strength & $\SI{100}{\pico\newton\,\micro\meter^{-2}}$ \\
    $\Gamma$ & Phase-field relaxation parameter & $\SI{0.5}{\micro\meter^{2}\,\pico\newton^{-1}\second^{-1}}$ \\
    $g$ & Cell-cell repulsion strength & $\SI{60}{\pico\newton}$ \\
    $\xi$ & Friction coefficient & $\SI{1500}{\pico\newton\,\second\,\micro\meter^{-2}}$  \\ [1ex] 
    \hline
    \end{tabular}
    \caption{Parameter values common for all simulations.}
    \label{tab:parameter_values}
\end{table}

\bibliography{main}

\section*{Acknowledgements}

This research is supported by the National Science Foundation (Award CMMI 1952912). The opinions, findings, and conclusions or recommendations expressed are those of the authors and do not necessarily reflect the views of the National Science Foundation.  This work uses the Bridges-2 system at the Pittsburgh Supercomputing Center (PSC) through allocation MCH220014 from the Advanced Cyberinfrastructure Coordination Ecosystem: Services and Support (ACCESS) program, which is supported by the National Science Foundation, grants 2138259, 2138286, 2138307, 2137603, and 2138296.

\section*{Data Availability}

Data sets generated during the current study will be made available from the corresponding author on reasonable request. 


\section*{Additional information}

\textbf{Competing interests} Authors declare no competing interest. 


\end{document}